\documentclass[aps,prb,twocolumn,showpacs,superscriptaddress,floatfix]{revtex4-1}

\usepackage[nottoc,numbib]{tocbibind}
\usepackage[colorlinks=true, citecolor=black, linkcolor=black, urlcolor=black]{hyperref}

\usepackage[all]{hypcap}
\usepackage{amsmath}
\usepackage{enumerate,bm}
\usepackage{graphicx}
\usepackage{grffile} 
\usepackage{color}
\usepackage{esint}
\usepackage{textpos}
\usepackage[usenames,dvipsnames]{xcolor}
\usepackage{subfigure}
\usepackage{flushend}
\usepackage{tabularx}

\usepackage{amssymb}
\usepackage{float}
\usepackage{subfigure}
\usepackage{ulem} 

\newcommand{\kk}{\vec{k}}
\newcommand{\QQ}{\vec{Q}}

\begin{document}

\author{M.~A.~Sentef}
\email[]{michael.sentef@mpsd.mpg.de}
\affiliation{Max Planck Institute for the Structure and Dynamics of Matter,
Center for Free Electron Laser Science, 22761 Hamburg, Germany}
\author{A.~Tokuno}
\affiliation{Centre de Physique Th\'eorique, 
\'Ecole Polytechnique, CNRS, 91128 Palaiseau Cedex, France}
\affiliation{Coll\`ege de France, 11 place Marcelin Berthelot, 75005 Paris, France}
\author{A.~Georges}
\affiliation{Centre de Physique Th\'eorique, 
\'Ecole Polytechnique, CNRS, 91128 Palaiseau Cedex, France}
\affiliation{Coll\`ege de France, 11 place Marcelin Berthelot, 75005 Paris, France}
\affiliation{Department of Quantum Matter Physics, University of Geneva, 24 Quai Ernest-Ansermet, 1211 Geneva 4, Switzerland}
\author{C.~Kollath}
\affiliation{HISKP, University of Bonn, Nussallee 14-16, D-53115 Bonn, Germany}


\title{
Theory of Laser-Controlled Competing Superconducting and Charge Orders
}
\date{\today}
\begin{abstract}
We investigate the nonequilibrium dynamics of competing coexisting superconducting (SC) and charge-density wave (CDW) orders in an attractive Hubbard model. A time-periodic laser field $\vec{A}(t)$ lifts the SC-CDW degeneracy, since the CDW couples linearly to the field ($\vec{A}$), whereas SC couples in second order ($\vec{A}^2$) due to gauge invariance. This leads to a striking resonance: When the photon energy is red-detuned compared to the equilibrium single-particle energy gap, CDW is enhanced and SC is suppressed, while this behavior is reversed for blue detuning. Both orders oscillate with an emergent slow frequency, which is controlled by the small amplitude of a third induced order, namely $\eta$ pairing, given by the commutator of the two primary orders. The \textit{induced} $\eta$ pairing is shown to control the enhancement and suppression of the dominant orders. Finally, we demonstrate that light-induced superconductivity is possible starting from a predominantly CDW initial state.
\end{abstract}
\maketitle

The nonequilibrium dynamics of solids stimulated by pump laser pulses and subsequently probed by various time-resolved spectroscopies has recently attracted lots of attention \cite{giannetti_ultrafast_2016, zhang_dynamics_2014}. In particular, nonequilibrium systems can host new states of matter that are not thermally accessible. Notable examples include nonthermal switching to hidden phases involving charge-density wave order \cite{stojchevska_ultrafast_2014} and Floquet-engineering of periodically driven band structures \cite{wang_observation_2013}. In particular, the prospect of controlling, enhancing, or possibly even inducing superconductivity (SC) with tailored light pulses \cite{fausti_light-induced_2011, kaiser_optically_2014, nicoletti_optically_2014, casandruc_wavelength-dependent_2015, mitrano_possible_2016} is tantalizing. Among the suggested mechanisms for light-induced superconductivity is the suppression of a competing order, such as a charge-density wave (CDW), in favor of superconductivity \cite{fausti_light-induced_2011, nicoletti_optically_2014, casandruc_wavelength-dependent_2015, patel_light-induced_2016}. The dynamics of ordered states with more than one order parameter were investigated theoretically previously \cite{akbari_theory_2013, moor_dynamics_2014, fu_quantum_2014, dzero_amplitude_2015, krull_coupling_2016} in different contexts.

Here we study a generic minimal model for competing SC and CDW orders with a focus on dynamically enhancing specifically one order by a tailored excitation. We consider the attractive Hubbard model on a 2D square lattice at half-filling, at which SC and CDW are degenerate due to $SO(4)$ symmetry \cite{yang_so4_1990}. The system is driven out of equilibrium by a classical homogeneous, time-dependent laser field that is included via Peierls substitution. The same form of driving via a classical field was used to predict a `Higgs' SC amplitude mode \cite{volkov_collisionless_1974, tsuji_theory_2015, cea_nonlinear_2016}, which was shown to be excited resonantly by THz pumping \cite{matsunaga_higgs_2013, matsunaga_light-induced_2014} and even nonresonantly by infrared pumping \cite{mansart_coupling_2013}. We use a mean-field approximation which takes into account both SC and CDW, and additionally $\eta$ pairing, i.e.~finite-momentum pairing at the CDW ordering wave vector\cite{yang_textit$eta$_1989}. This assures that the $SO(4)$ symmetry is preserved. The ensuing nonlinear coupled differential equations with a self-consistency condition are solved numerically, starting from a coexisting state with equal SC and CDW at equilibrium or a predominant CDW state, respectively. We find a resonance effect when the photon frequency $\omega$ is of the order of the single-particle energy gap $2 \Delta_0$. CDW is favored for red detuning ($\omega < 2 \Delta_0$), while blue detuning ($\omega > 2 \Delta_0$) favors SC. Importantly, a finite expectation value of $\eta$ pairing is found to be induced and to control the SC and CDW dynamics. 

We investigate the fermionic 2D square-lattice attractive Hubbard model at half-filling,
\begin{align}
H &= \sum_{\kk \sigma} \epsilon_{\kk} n_{\kk\sigma} + U \sum_i n_{i\uparrow} n_{i\downarrow}
\end{align}
with single-particle energy dispersion $\epsilon_{\kk} = - 2 J (\cos(k_x)+\cos(k_y))$. Here $J$ is the nearest-neighbor hopping, $\kk = (k_x, k_y) \in (-\pi,\pi] \times (-\pi,\pi]$ are dimensionless momenta, $n_{\kk\sigma} = c^{\dagger}_{\kk\sigma} c^{}_{\kk\sigma}$ is the number operator with fermionic annihilation (creation) operators $c^{(\dagger)}_{\kk\sigma}$, and $U$ is the onsite interaction. We choose $J=$ 0.25 eV and the attraction $U=-0.2188$ eV. The interaction term is mean-field decoupled focusing on the relevant SC, CDW, and $\eta$ pairing instabilities for $U<0$ at half-filling,
\begin{align}
&f_{\kk} \equiv \langle c^{}_{-\kk\downarrow} c^{}_{\kk\uparrow}\rangle, g_{\kk} \equiv \frac12 \sum_{\sigma} \langle c^{\dagger}_{\kk\sigma} c^{}_{\kk+\QQ\sigma}\rangle, \eta_{\kk} \equiv \langle c^{}_{-(\kk+\QQ)\downarrow} c^{}_{\kk\uparrow} \rangle, \nonumber \\
&\Delta_{SC} \equiv U \sum_{\kk} f_{\kk}, \Delta_{CDW} \equiv U \sum_{\kk} g_{\kk}, \Delta_{\eta} \equiv U \sum_{\kk} \eta_{\kk},
\label{eq:selfconsistency}
\end{align}
leading to
\begin{align}
&H_{MF} = \sum_{\kk} \Psi^{\dagger}_{\kk} h_{\kk} \Psi_{\kk}, \nonumber \\
&h_{\kk} \equiv 
\left( \begin{array}{cccc} \epsilon_{\kk-\vec{A}} & \Delta^*_{CDW} & \Delta_{SC} & \Delta_{\eta} \\ \Delta_{CDW} & \epsilon_{\kk-\vec{A}+\QQ} & \Delta_{\eta} & \Delta_{SC} \\ \Delta^*_{SC} & \Delta^*_{\eta}  & -\epsilon_{-(\kk+\vec{A})} & -\Delta_{CDW} \\ \Delta^*_{\eta}  & \Delta^*_{SC} & -\Delta^*_{CDW} & -\epsilon_{-(\kk+\vec{A}-\QQ)} \end{array} \right),
\end{align}
with spinors $\Psi^{\dagger}_{\kk} \equiv (c^{\dagger}_{\kk\uparrow}, c^{\dagger}_{\kk+\QQ\uparrow}, c^{}_{-\kk\downarrow}, c^{}_{-(\kk+\QQ)\downarrow} )$, and $\QQ=(\pi,\pi)$ is the CDW ordering wave vector. A very similar model was investigated to identify Raman signatures of the Higgs mode in systems with coexisting SC and CDW orders \cite{cea_nature_2014}. We note that the inclusion of $\eta$ pairing is necessary to close the $SO(4)$ algebra. More generally, a third order is induced whenever there is a dynamical competition between two non-commuting orders, as noted in Ref.~\onlinecite{das_nodeless_2013}. The system is driven out of equilibrium by a time-dependent laser field $\vec{A}(t)$, measured in the same dimensionless units as the momenta, with electric field $\vec{E}(t) = -\partial_t \vec{A}(t)$, included via Peierls substitution $\epsilon_{\kk} \rightarrow \epsilon_{\kk-\vec{A}}$. 

The Heisenberg equations of motion (EOMs) for the momentum expectation values are found as   
\begin{align}
i \partial_t n_{\kk} =& - \Delta_{SC} (f_{\kk} - f^*_{\kk}) + \Delta_{CDW} (g_{\kk} - g^*_{\kk}) \nonumber \\ & -\Delta^*_{\eta} \eta_{\kk} + \Delta_{\eta} \eta^*_{\kk},
\nonumber \\
i \partial_t f_{\kk} =& 
\Delta_{SC} (1 - (n_{\kk} + n_{-\kk})) + (\epsilon_{\kk-\vec{A}} + \epsilon_{\kk+\vec{A}}) f_{\kk} \nonumber \\ & + \Delta_{CDW} (\eta_{\kk} + \eta_{\kk+\QQ}) - \Delta_{\eta} (g^*_{\kk} + g^*_{-\kk}), 
\nonumber \\
i \partial_t g_{\kk} =& \Delta_{CDW} (n_{\kk} - n_{\kk+\QQ}) - 2 \epsilon_{\kk-\vec{A}} g_{\kk} \nonumber \\ & + \Delta_{SC} (\eta^*_{\kk} - \eta_{\kk+\QQ}) + \Delta_{\eta} f^*_{\kk} - \Delta^*_{\eta} f_{\kk+\QQ}, 
\nonumber \\
i \partial_t \eta_{\kk} =& (\epsilon_{\kk-\vec{A}} - \epsilon_{\kk+\vec{A}}) \eta_{\kk} + \Delta_{CDW} (f_{\kk} + f_{\kk+\QQ}) \nonumber \\ & - \Delta_{SC} (g_{-\kk}+g^*_{\kk}) - \Delta_{\eta} (n_{\kk} + n_{-(\kk+\QQ)} - 1),
\label{eq:eoms}
\end{align}
where we suppress time arguments for brevity, set $\hbar = 1$, and $n_{\kk} \equiv \frac12 \sum_{\sigma} \langle c^{\dagger}_{\kk\sigma} c^{}_{\kk\sigma} \rangle$ is the momentum occupation per spin. These equations are solved on a grid with 120 $\times$ 120 momentum points using time-ordered exponentials with a fourth-order commutator-free scheme \cite{alvermann_high-order_2011}, and independently checked with fourth-order Runge-Kutta integration, together with instantaneous self-consistency conditions for the $\Delta$'s according to Eq.~(\ref{eq:selfconsistency}). Convergence in the time step size was checked; for the former a time step of 0.1 $\hbar/\text{eV} \approx 0.066 \; \text{fs}$  was found to be sufficient. The EOMs in Eq.~(\ref{eq:eoms}) are initialized with equilibrium self-consistent solutions, which for our choice of parameters at zero temperature are given by $\Delta_{0} = \sqrt{\Delta^2_{SC,0} + \Delta^2_{CDW,0}}=$ 0.01 eV. The equilibrium single-particle energy gap is $2 \Delta_0$. The $\eta$ pairing is initially zero, $\Delta_{\eta,0} = 0$. 

Importantly the laser field breaks the degeneracy between SC and CDW, as can be seen by expanding the field-dependent terms on the right-hand sides of Eq.~(\ref{eq:eoms}) in a small $\vec{A}$. For the CDW one has $2 \epsilon_{\kk-\vec{A}} = 2 \epsilon_{\kk} - 2 \vec{v}_{\kk} \vec{A} +\mathcal{O}(\vec{A}^2)$, with band velocity $\vec{v}_{\kk} \equiv \partial_{\kk} \epsilon_{\kk}$. By contrast, for the SC one obtains $\epsilon_{\kk-\vec{A}} + \epsilon_{\kk+\vec{A}} = 2 \epsilon_{\kk} + \mathcal{O}(\vec{A}^2)$, which does not contain a linear term in the field\cite{tsuji_theory_2015}. This difference is due to the fact that photons directly couple to the charge modulation of the CDW, whereas such a linear coupling is forbidden for the SC due to gauge invariance. Importantly, the actual dynamics of the momentum-integrated order parameters is only affected directly to $\mathcal{O}(\vec{A}^2)$, but the difference in coupling to the momentum-resolved anomalous expectation values turns out to be crucial for the results. 

In the following, we choose a linearly polarized continuous wave laser excitation with $A_x(t) = A_y(t) = A_{\text{max}} \sin(\omega t)$, with a fixed small amplitude $A_{\text{max}} = 5 \times 10^{-5}$ in dimensionless units, which corresponds to a peak electric field strength $E_{\text{max}}[\text{V/\AA}] = \sqrt{2} \times \omega[\text{eV}] \times A_{\text{max}} / a[\text{\AA}]$, where $a$ is the lattice constant. For example, $a = 2$ $\text{\AA}$ and $\omega = 0.01$ eV implies $E_{\text{max}} = 3.5 \times 10^{-7} \text{V/\AA} = 35$ V/cm.

\begin{figure}[ht!pb]
\includegraphics[clip=true, trim=0 0 0 0,width=\columnwidth]{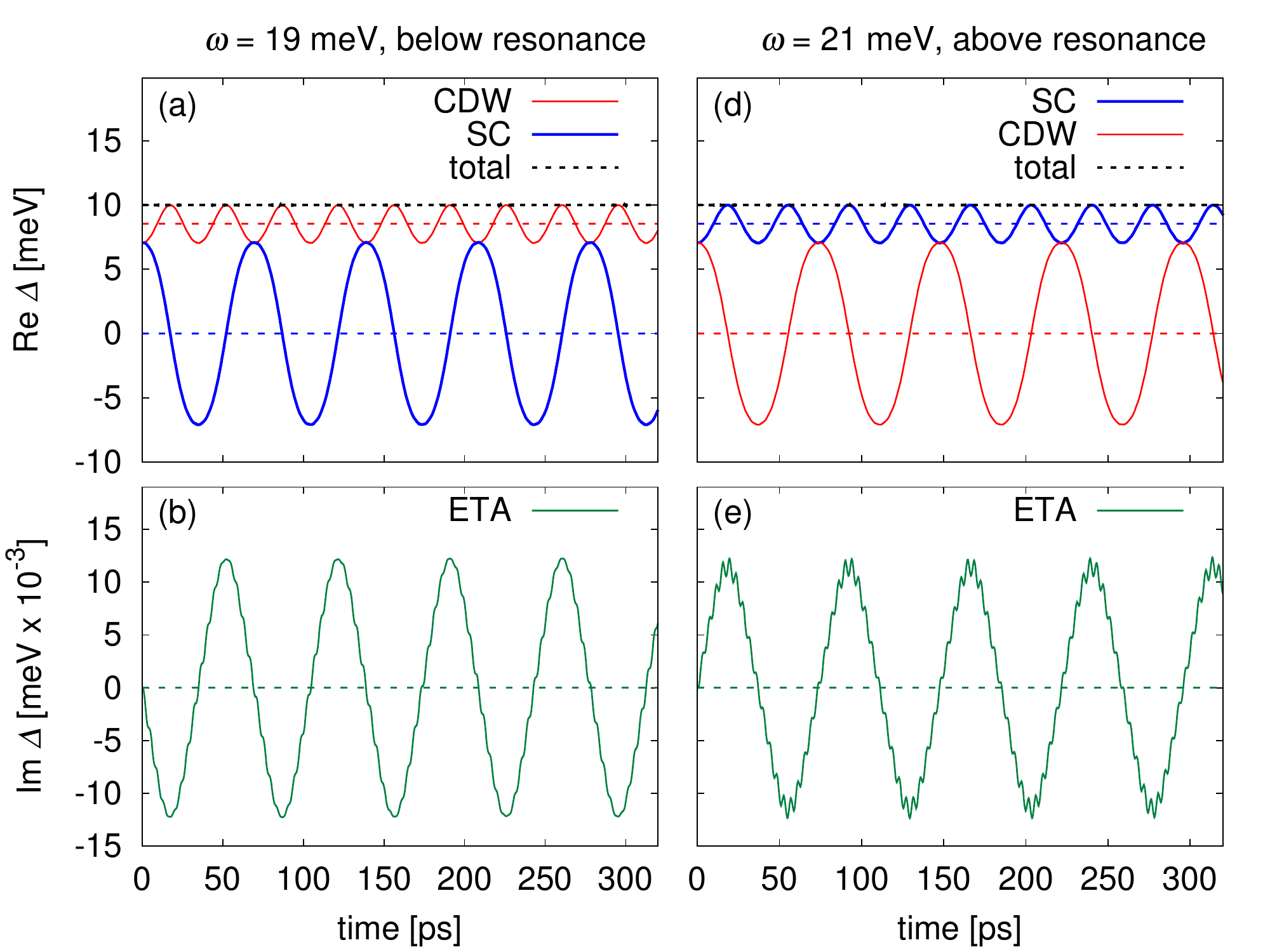}
\includegraphics[clip=true, trim=0 0 0 0,width=\columnwidth]{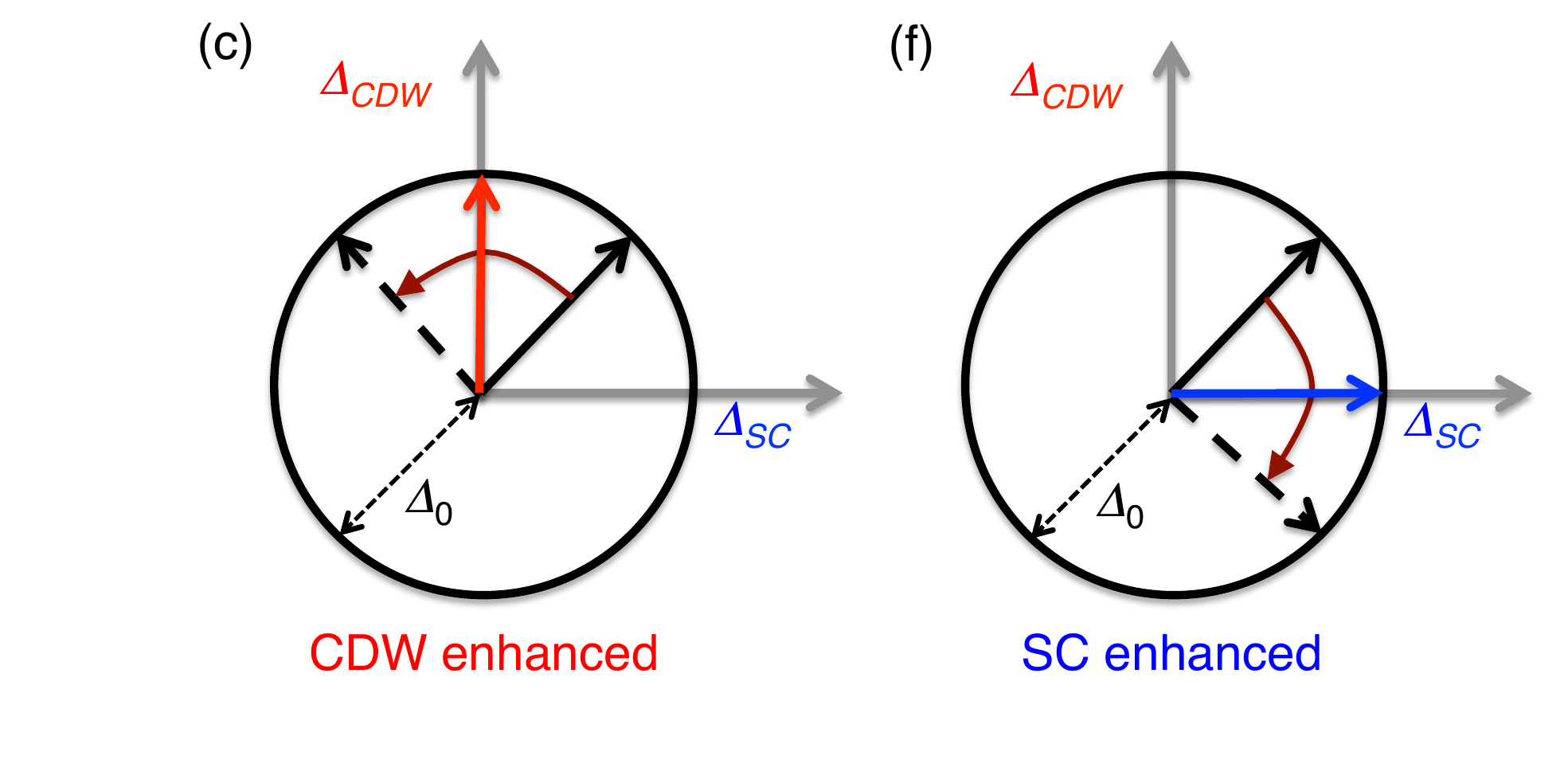}
\vspace{-1.0cm}
\caption{
{\bf Laser-controlled order.} 
(a) $\Delta_{SC}(t)$ and $\Delta_{CDW}(t)$, and total half-gap $\sqrt{\Delta^2_{SC}(t) + \Delta^2_{CDW}(t)}$, for a driving field with $\omega=$ 19 meV (red-detuned from 2 $\Delta_0$). (b) The corresponding $\Delta_{\eta}(t)$ (``ETA''). (c) Depiction of the dynamics in the $\Delta_{SC}$-$\Delta_{CDW}$ plane with enhanced CDW. (d), (e), (f) The same for a driving field with $\omega=$ 21 meV (blue-detuned from 2 $\Delta_0$) with enhanced SC. In all cases, dashed colored lines show the respective time averages.}
\label{fig1}
\end{figure}

We first choose an initial state with $\Delta_{SC,0} = \Delta_{CDW,0} = \Delta_{0}/\sqrt{2}$ and vary the driving frequency $\omega$ at fixed $A_{\text{max}}$. The most striking effect is found near $\omega = 2 \Delta_0 = 20$ meV, see Fig.~\ref{fig1}, which is different from the Anderson pseudospin resonance at $2 \omega = 2 \Delta_0$ for the SC-only case \cite{tsuji_theory_2015}. For red detuning, $\omega < 2 \Delta_0$, we find an enhancement of time-averaged CDW and a suppression of time-averaged SC. The time-dependent order parameters show regular oscillations (Fig.~\ref{fig1}(a)). At the same time, a nonzero $\Delta_{\eta}$ is induced (Fig.~\ref{fig1}(b)) and found to oscillate around zero. $\Delta_{\eta}$ is imaginary in the gauge where $\Delta_{SC}$ and $\Delta_{CDW}$ are real. For our choice of parameters, $\Delta_\eta$ is three orders of magnitude smaller than the other order parameters, yet plays a crucial role for the competing order dynamics. 

First, we observe that a very slow time scale emerges for the entire order parameter dynamics. $\Delta_{\eta}$ oscillates at the same frequency $\omega_{\text{slow}}$ as the slowly oscillating $\Delta_{SC}$. Note that $2 \Delta_0$ corresponds to an oscillation period of 0.03 ps, whereas $\omega_{\text{slow}}$ corresponds to a much longer one of 70 ps. The oscillation frequency of the light-enhanced order is $2 \omega_{\text{slow}}$. 

We find that the total half-gap $\sqrt{\Delta^2_{SC}(t) + \Delta^2_{CDW}(t)}$ remains almost constant $= \Delta_0$ over time (see Fig.~\ref{fig1}(a),(d)), with relative deviations of order $10^{-4}$ for our choice of driving field. This approximate conservation law then simply explains the frequency doubling for the enhanced order by a composite vector order parameter of fixed length that oscillates in the $\Delta_{SC}$-$\Delta_{CDW}$ plane, see Fig.~\ref{fig1}(c). Obviously, if for example $\Delta_{SC}(t) = \frac{\Delta_0}{\sqrt{2}} \cos(\omega_{\text{slow}} t)$, then $\Delta_{CDW}(t) = \Delta_0 \sqrt{\frac34 - \frac14 \cos(2\omega_{\text{slow}} t)}$.

In the next step, we increase the laser frequency to the blue-detuned case, $\omega > 2 \Delta_0$. Here we find the exact opposite behavior than for the red-detuned case: $\Delta_{SC}$ is enhanced and $\Delta_{CDW}$ is suppressed (Fig.~\ref{fig1}(d)). Simultaneously, $\text{Im} \Delta_{\eta}$ reverses its sign (Fig.~\ref{fig1}(e)) and is initially positive, coinciding with the enhancement of $\Delta_{SC}$ (Fig.~\ref{fig1}(f)), whereas the initally negative $\text{Im} \Delta_{\eta}$ in Fig.~\ref{fig1}(b) coincided with CDW enhancement for the red-detuned case. 

\begin{figure}[ht!pb]
\includegraphics[clip=true, trim=0 0 0 0,width=0.9\columnwidth]{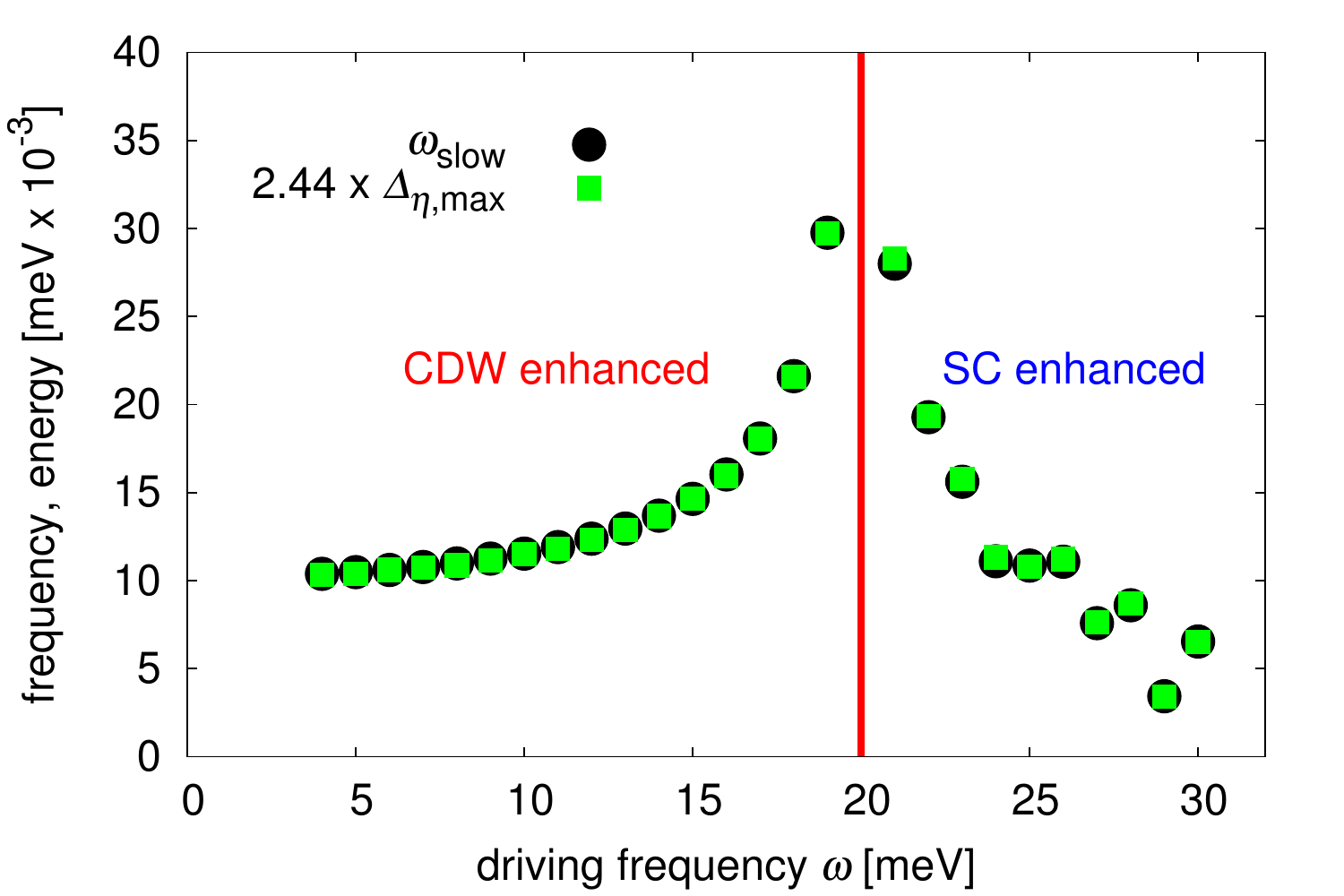}
\caption{
{\bf Nonequilibrium orders controlled by gap resonance.} 
Oscillation frequency $\omega_{\text{slow}}$ (black circles), obtained from the time-dependent $\Delta$'s, as well as amplitude $\Delta_{\eta,\text{max}}$ of the time-dependent $\Delta_{\eta}$, plotted as a function of the driving frequency $\omega$. The red vertical line indicates the $\omega = 2 \Delta_0$ resonance.}
\label{fig2}
\end{figure}

Having noted the important role of $\eta$ pairing, we now turn to the systematics of the nonequilibrium dynamics as a function of the driving frequency $\omega$. First we notice that in the range of frequencies below and above the $2 \Delta_0$ resonance shown in Fig.~\ref{fig2}, $\Delta_{CDW}$ is always enhanced below the resonance and $\Delta_{SC}$ is enhanced above the resonance. To understand the origin of the very small energy scale setting the oscillation frequencies, we show in Fig.~\ref{fig2} the dependence of the observed oscillation frequency $\omega_{\text{slow}}$ and of the amplitude $\Delta_{\eta,\text{max}}$ of the $\eta$ pairing oscillations on the driving frequency. Empirically we find a ratio $\Delta_{\eta,\text{max}}/\omega_{\text{slow}}$ = 2.44 $\pm$ 0.02 independent of driving frequency for the data points in Fig.~\ref{fig2} \footnote{We note that the observed ratio of 2.44 in this case is not universal, but indeed depends on the specific choice of parameters, for instance on the ratio between $\Delta_0$ and the electronic bare bandwidth $4 J$.}, with strongly increased values when approaching the resonance \footnote{Exactly at the resonance, very irregular behavior is indeed found.}. It is evident that the induced $\eta$ pairing not only determines via its sign the enhancement or suppression of SC, but at the same time sets the slow oscillation frequency of the other order parameters.

The central result of this work is the possibility to enhance either SC or CDW order above and below the $2 \Delta_0$ resonance. Which of the orders is enhanced depends on the initial sign of the imaginary part of $\Delta_{\eta}$, independent of the exact choice of parameters. In order to gain some analytical understanding of the change in enhancement and suppression above and below the resonance, we take a closer look at the early-time dynamics by keeping the $\Delta$'s on the right-hand side of the EOMs fixed, using $\Delta_{SC} = \Delta_{CDW} = \Delta_{0}/\sqrt{2}$, and linearizing in the field $\vec{A}(t)$. This amounts to solving the equations
\begin{align}
i \partial_t \delta n_{\kk} =& - \frac{\Delta_{0}}{\sqrt{2}} \delta(f_{\kk} - f^*_{\kk}) + \frac{\Delta_{0}}{\sqrt{2}}\delta(g_{\kk} - g^*_{\kk}) 
\nonumber \\
i \partial_t \delta f_{\kk} =& 
\frac{\Delta_{0}}{\sqrt{2}} \delta(1 - (n_{\kk} + n_{-\kk})) + \frac{\Delta_{0}}{\sqrt{2}} (\eta_{\kk} + \eta_{\kk+\QQ}), 
\nonumber \\
i \partial_t \delta g_{\kk} =& \frac{\Delta_{0}}{\sqrt{2}} \delta (n_{\kk} - n_{\kk+\QQ}) - 2 \epsilon_{\kk} \delta g_{\kk} + 2 \vec{v}_{\kk} \vec{A} g_{\kk,0} \nonumber \\ & + \frac{\Delta_{0}}{\sqrt{2}} (\eta^*_{\kk} - \eta_{\kk+\QQ}), 
\nonumber \\
i \partial_t \eta_{\kk} =& \frac{\Delta_{0}}{\sqrt{2}} \delta (f_{\kk} + f_{\kk+\QQ}) - \frac{\Delta_{0}}{\sqrt{2}} \delta (g_{-\kk}+g^*_{\kk}),
\end{align}
where $g_{\kk,0} = -\frac{\Delta_{0}}{\sqrt{2} E_{\kk}}$, $E_{\kk} \equiv \sqrt{\epsilon_{\kk}^2 + \Delta_0^2}$, and $\delta f_{\kk}(t) \equiv f_{\kk}(t)-f_{\kk}(0)$ etc. These equations can be solved via Laplace transforms and in particular yield for the induced $\eta$ pairing to lowest order
\begin{equation}
\eta_{\kk,1}(t) = -A_{\kk,0} \Delta_0 g_{\kk,0} \frac{-\omega \sin(2 E_{\kk} t) + 2 E_{\kk} \sin(\omega t)}{
 E_{\kk} (4 E_{\kk}^2 - \omega^2)},
\end{equation}
with $A_{\kk,0} \equiv A_{\text{max}} (v_{\kk,x} + v_{\kk,y})$. The vanishing imaginary part of $\eta_{\kk,1}$ together with the odd-in-momentum real part due to $A_{-\kk,0} = - A_{\kk,0}$, implies that $\Delta_{\eta,1}(t) = 0$. However, if we use $\eta_{\kk,1}$ as a seed for the next iteration, focusing on the next order in the field of the imaginary part of $\eta$ pairing, we find
\begin{align}
\text{Im} \; \eta_{\kk,2}(t) &= 2 A_{\kk,0} \int_0^t \eta_{\kk,1}(t') \sin(\omega t') dt',  \nonumber \\
&= \frac{2 A_{\kk,0}^2 \Delta_0 g_{\kk,0} t}{4 E_{\kk}^2 - \omega^2} + \eta_{\kk,2,\text{osc}}(t),
\end{align}
where we isolate the first term, which grows linearly in time. The remaining terms $\eta_{\kk,2,\text{osc}}(t)$ oscillate with frequency $\omega$ and time-average to zero.

Noting that the dominant contribution comes from near the Fermi level, where $\epsilon_{\kk} = 0$ and $E_{\kk} = \Delta_0$, this result explains the $\omega = 2 \Delta_0$ resonance and shows how the laser frequency controls the initial sign of the \textit{induced} $\Delta_\eta$. Importantly, below the resonance $\text{Im} \; \eta_{\kk,2}$ is positive, hence $\Delta_{\eta}$ is negative, with a sign change when going above resonance, as observed in the numerics. Together with the correlation between this sign and the respective upturn or downturn of $\Delta_{SC}$ and $\Delta_{CDW}$ (see Fig.~\ref{fig1}), the laser control of SC and CDW orders is thus understood as a consequence of the linear-in-the field coupling of charge-modulated orders versus the quadratic-in-the-field coupling of the superconducting condensate, together with the way SC and CDW orders couple to $\eta$ pairing in Eq.~\ref{eq:eoms}. Notice that this coupling is generic: $\eta$ pairing is given by the commutator between the SC and CDW operators, whose expectation values determine the gap values according to Eq.~\ref{eq:selfconsistency}. Therefore the mathematical structure enabling the induced $\eta$ pairing to control the enhancement and suppression of SC and CDW appears naturally for competing orders.

\begin{figure}[ht!pb]
\includegraphics[clip=true, trim=0 0 0 0,width=\columnwidth]{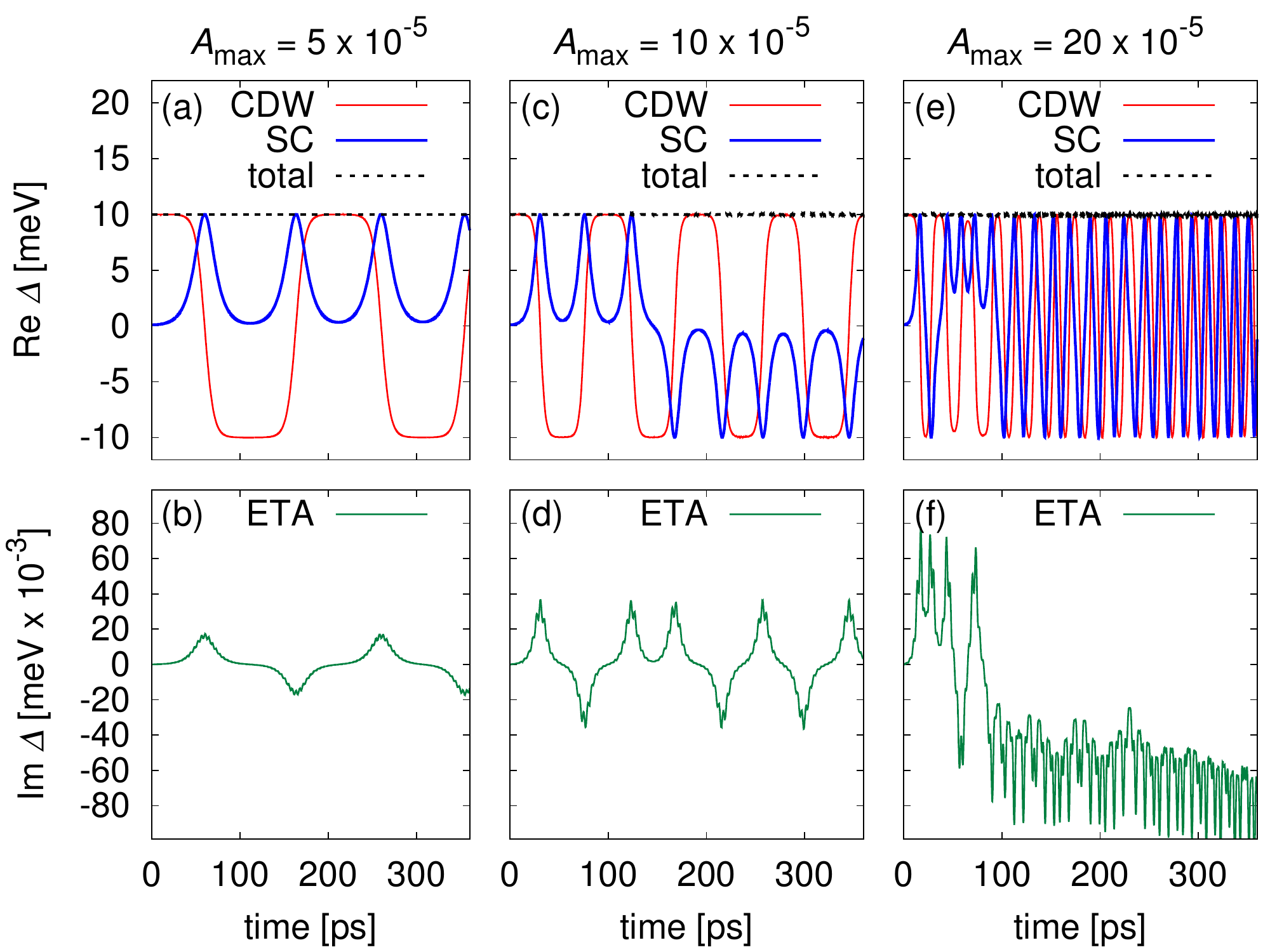}
\caption{
{\bf Light-induced superconductivity.} 
(a) $\Delta_{SC}(t)$ and $\Delta_{CDW}(t)$ for a driving field with $\omega=$ 21 meV and $A_{\text{max}} = 5 \times 10^{-5}$, starting from an initial state with mostly CDW order. (b) The corresponding $\Delta_{\eta}(t)$ (``ETA''). (c), (d) The same for $A_{\text{max}} = 10 \times 10^{-5}$. (e), (f) The same for $A_{\text{max}} = 20 \times 10^{-5}$. Dashed black line indicates the total half-gap.}
\label{fig3}
\end{figure}

Finally, we turn to the question as to whether this mechanism can also explain light-induced superconductivity when starting from an initial state with predominant CDW order. To this end, we investigate the case in which we choose an initial solution with $\frac{\Delta_{CDW,0}}{\Delta_{SC,0}} = 99$. This ratio is chosen to provide a seed for $\Delta_{SC}$ which is needed in a mean-field treatment to obtain a nonzero $\Delta_{SC}$. We show the dynamics for blue-detuned driving fields with three different maximal field strengths in Fig.~\ref{fig3}. Apparently it is possible to light-induce SC starting from a state which has predominant CDW order. The approximate conservation of the total gap is still observed. Thus in all cases the maximal SC order reached corresponds to the initial CDW order. At small field strength, a regular oscillation is found for the considered times, whereas at larger driving fields the sign of the SC order can change and regular oscillations are only seen in certain time windows. The regular oscillations behave very similarly to the previously considered case of a balanced initial order. In particular, a finite value of $\text{Im} \Delta_{\eta}$ is again induced. Its oscillation frequency corresponds to the one of the CDW order, and the induced SC order has twice this frequency. As in the case of the initially balanced order, the slow oscillation frequency in the regular part of the oscillations corresponds again to the amplitude of the induced $\eta$ pairing. The time on which the initial switching from CDW to SC happens, i.e.~the time for SC to reach its first maximum, scales approximately inversely with the field strength $A_{\text{max}}$. This can be seen from Fig.~\ref{fig3} by noting that the first maximum of $\Delta_{SC}$ is reached in half the time when $A_{\text{max}}$ is doubled, as is the amplitude of $\eta$ pairing. Notice that this observation is again consistent with the fact that the oscillation frequency scales linearly with the induced $\eta$ pairing. In addition, we note that we have also checked that light-induced superconductivity is stable after the field is switched off in a situation with a laser pulse of finite duration. In that case, $\eta$ pairing is induced and remains constant after the pulse, while $\Delta_{SC}$ and $\Delta_{CDW}$ continue oscillating, preserving the total gap, at a slow frequency determined by the magnitude of $\Delta_{\eta}$.

In conclusion, we solved a minimal model of competing coexisting orders in the time domain. A continuous-wave laser tuned to frequencies near the $2 \Delta_0$ resonance was shown to control the orders in real time on picosecond time scales for extremely small laser intensities. This low-field stimulation of coexisting orders apparently requires a symmetry between these orders, in this case $SO(4)$ symmetry, leading to a perfect ground-state degeneracy and the existence of a long-wavelength Goldstone mode that corresponds to a rotation of the general vector order parameter. If this degeneracy did not exist, it would cost a finite amount of excitation energy to rotate from one state to the other. Importantly, $SO(4)$ symmetry is an exact symmetry of the studied model and not an artefact of the employed mean-field approximation.

Competing superconductivity and density-wave orders appear in a host of materials ranging from cuprates \cite{tranquada_evidence_1995, ghiringhelli_long-range_2012, chang_direct_2012, hayward_angular_2014, cyr-choiniere_suppression_2015} via Fe-based superconductors \cite{de_la_cruz_magnetic_2008, chubukov_renormalization_2009} to $2H$-NbSe$_2$\cite{wilson_charge-density_1974, yokoya_fermi_2001, bawden_spin-valley_2016} or bismuthates\cite{uchida_superconductivity_1987, giraldo-gallo_field-tuned_2012}. Also correlated heterostructures provide an additional playground for competing CDW and SC orders\cite{frano_long-range_2016}. Future theoretical work should address situations with only nearly degenerate competing orders.  Moreover, the role of strong correlation effects beyond the mean-field approximation, which may cause relaxation of order parameter dynamics\cite{tsuji_nonthermal_2013}, should be investigated. Additionally, the role of dissipation, either by adding a phenomenological damping term in the equations of motion, or more realistically by including electron-phonon scattering, which was suggested to also play a role for light-enhanced superconductivity in Ref.~\onlinecite{sentef_theory_2016}, should be studied. 

\textit{Acknowledgment.--} We acknowledge stimulating discussions with A.~Cavalleri and A.~Kampf. This work was financially supported by the DFG through the Collaborative Research Center 1238 project C05 (C.K.) and through the Emmy Noether programme (M.A.S.). We further acknowledge financial support by the European Research Council through ERC-319286 QMAC (A.G.) and ERC-648166 Phonton (C.K.). 

\bibliography{SC-CDW}


\end{document}